\documentclass[12pt]{article}
\usepackage{amsmath}
\begin{document}

\title{The $n$-level spectral correlations \\ for
chaotic systems}

\author{Taro Nagao$^1$ and Sebastian M{\"u}ller$^2$}

\date{}
\maketitle

\begin{center}
\it $^1$ Graduate School of Mathematics, Nagoya University, Chikusa-ku, \\ Nagoya 464-8602, Japan \\
\it $^2$ Department of Mathematics, University of Bristol, Bristol BS8 1TW, UK
\end{center}

\begin{abstract}

We study the $n$-level spectral correlation functions of classically chaotic quantum systems without
time-reversal symmetry. According to Bohigas, Giannoni and Schmit's universality conjecture, it is
expected that the correlation functions are in agreement with the prediction of the Circular Unitary
Ensemble (CUE) of random matrices. A semiclassical resummation formalism allows us to express the
correlation functions as sums over pseudo-orbits. Using an extended version of the diagonal
approximation on the pseudo-orbit sums, we derive the $n$-level correlation functions identical to
the $n \times n$ determinantal correlation functions of the CUE.

\end{abstract}

PACS: 05.45.Mt; 02.50.-r

\medskip

KEYWORDS: quantum chaos; periodic orbit theory; random matrices

\newpage

\section{Introduction}
\setcounter{equation}{0}
\renewcommand{\theequation}{1.\arabic{equation}}

Quantum systems whose classical limit is chaotic display universal spectral statistics. Their
spectral correlations depend only on the symmetry class of the system, and agree with predictions
obtained from averaging over ensembles of random matrices\cite{BGS,MK,CVG,BERRY87}. For instance, 
let us suppose that the time-reversal invariance of a chaotic system is broken by applying a 
magnetic field. Then, in the semiclassical limit, it is conjectured that its spectral correlation 
functions are in agreement with the predictions of the Circular Unitary Ensemble (CUE) (or 
the Gaussian Unitary Ensemble (GUE)) of random matrices.
\par
A way to understand the origins for this universality is provided by semiclassics. In the semiclassical 
theory, the $2$-level spectral correlation function is expressed as a sum over the pairs of periodic 
orbits. Berry introduced a useful scheme called the
diagonal approximation\cite{BERRY85}. In this scheme, when the time-reversal invariance is broken, only
the sum over the pairs of identical periodic orbits is taken into account. Then one can derive the
smooth (non-oscillatory) part of the correlation function in agreement with the CUE. Shukla extended
this scheme to calculate the $n$-level spectral correlation functions and succeeded in deriving
asymptotic forms of the Fourier transforms in agreement with the CUE\cite{PS}.
\par
The diagonal approximation brought about a great progress in understanding universality. However,
it is able to only partially reproduce the random matrix predictions, even if we restrict ourselves
to the case of broken time-reversal symmetry. It is an approximation which yields only the smooth
parts of the correlation functions and misses the remaining oscillatory parts.
\par
Refined schemes to reproduce the full predictions have been developed in the study of the
Riemann zeta function. It is conjectured that the complex zeros of the Riemann zeta function are
mutually correlated in a similar way as the energy levels of chaotic quantum systems without
time-reversal symmetry\cite{MONT,DAH,RS94,RS96}. Correlation functions of the zeros can be 
written as the multiple sums over prime numbers similar to the periodic orbit sums in semiclassics. 
One is thus able to develop a scheme analogous to the semiclassical periodic orbit theory. 
Using additional input from number theory, it is then possible to access oscillatory 
contributions as well.  As a result, the full CUE correlation functions have been reproduced 
under certain assumptions in several works\cite{K93,BK95,BK96,CS06,CS08}.
\par
In the case of chaotic quantum systems, although the problem is in some respect more involved, the
analogous question was addressed in \cite{BDA,H07,KM}. Following progress in the method of
semiclassical diagrammatic expansions\cite{SR,M04,M05,SM}, Heusler et al. proposed a way to
evaluate semiclassically both oscillatory and smooth parts of the full $2$-level correlation
function, and obtained results agreeing with the random matrix prediction\cite{H07}. 
Keating and M\"uller recently gave a justification of Heusler et al.'s argument\cite{KM}. 
The essential idea is to relate the correlation function to a generating
function (a ratio of spectral determinants) and then make use of an improved (``resummed'')
semiclassical approximation for the latter, the so-called Riemann-Siegel lookalike formula
established by Berry and Keating\cite{BK00,K02,BK02}. For systems without time-reversal 
invariance, a generalization of the diagonal approximation to this setting is sufficient 
to derive the full $2$-level correlation function.
\par
In this paper, we apply a generalization of Keating and M\"uller's method to calculate the $n$-level
correlation functions of chaotic quantum systems without time-reversal symmetry. In \S 2, we develop
a generating function formalism for the $n$-level correlation functions, introduce the Riemann-Siegel
lookalike formula, and put it into the formalism. In \S 3, an extended version of the diagonal
approximation is formulated and the $n$-level correlation functions are calculated in the
semiclassical limit. The resulting formulas involve sums over several different contributions 
generalizing the sum over the smooth and oscillatory parts for the $2$-level correlation function.
In \S 4, we verify that these sums are identical to the $n \times n$ determinantal formulas of the 
CUE correlation functions known from random-matrix theory (RMT). This is done by establishing 
their agreement with a representation of the random-matrix average obtained by Conrey and 
Snaith\cite{CS08} (interestingly, this representation had originally been developed to facilitate 
the comparison to number-theoretic rather than semiclassical results). We thus confirm that 
the known partial results based on the diagonal approximation are extended to the forms 
in agreement with the full random matrix predictions. The last section is devoted to a brief
discussion on the result.

\section{Generating function}
\setcounter{equation}{0}
\renewcommand{\theequation}{2.\arabic{equation}}

Let us suppose that $H$ denotes the Hamiltonian of a bounded quantum system which is chaotic in the
classical limit. We are interested in the distribution of the energy levels $E_j$ (the eigenvalues of
$H$). The density of these energy levels
\begin{equation}
\rho(E) = \sum_j \delta(E - E_j)
\end{equation}
may be separated into the smoothed part
\begin{equation}
\label{rhoE} {\bar \rho}(E) \sim \frac{\Omega(E)}{(2 \pi \hbar)^f}
\end{equation}
and the fluctuation around it. Here $\Omega(E)$ is the volume of the energy shell in the classical
phase space and $f > 1$ is the number of degrees of freedom.
\par
We now want to determine the $n$-level correlation functions, which are defined as
\begin{equation}
\label{Rdef} R_n(\epsilon_1,\cdots,\epsilon_n) = \frac{1}{{\bar \rho}^n} \left\langle \prod_{j=1}^n \rho(E
+ \epsilon_j) \right\rangle
\end{equation}
and describe the fluctuation of the energy level distribution around the smoothed density ${\bar
\rho}$. To obtain a smooth function, we take the average $\langle \cdot \rangle$ over 
the windows of the center energy $E$ and energy differences $\epsilon_j$. 
\par
The idea of the generating function formalism is to represent the level densities in (\ref{Rdef}) through
traces of the resolvent,
\begin{equation}
\rho(E) = \frac{i}{2 \pi} \left( {\rm Tr}\frac{1}{E^+ - H} - {\rm Tr}\frac{1}{E^- - H} \right)
\end{equation}
(where $E^{\pm} = E \pm i \kappa$, and $\kappa$ is an infinitesimal positive number), and then
express these traces in terms of derivatives of the spectral determinant $\Delta(E) = \det(E - H)$,
\begin{equation}
{\rm Tr}\frac{1}{E - H}=- \left. \frac{\partial}{\partial
\epsilon}\frac{\Delta(E)}{\Delta(E+\epsilon)} \right|_{\epsilon=0}.
\end{equation}
Eq.(\ref{Rdef}) then turns into
\begin{eqnarray}
\label{RnZn} R_n(\epsilon_1,\cdots,\epsilon_n) & = & \left( \frac{i}{2 \pi {\bar \rho}} \right)^n
\left\langle \prod_{j=1}^n \left\{ \sum_{\sigma_j = \pm 1} \sigma_j {\rm Tr}\frac{1}{ E + \epsilon_j
+ i \sigma_j \kappa_j - H} \right\} \right\rangle \nonumber \\ & = & \left.
\frac{\partial^n}{\partial \epsilon_1 \partial \epsilon_2 \cdots
\partial \epsilon_n}
Z_n \right|_{\boldsymbol\eta = \boldsymbol\epsilon},
\end{eqnarray}
where
\begin{equation}
\boldsymbol\epsilon = (\epsilon_1,\epsilon_2,\cdots,\epsilon_n), \ \ \ \boldsymbol\eta =
(\eta_1,\eta_2,\cdots,\eta_n)
\end{equation}
and the generating function $Z_n$ is defined as
\begin{equation}
\label{Zn} Z_n = \left( \frac{1}{2 \pi {\bar \rho} i} \right)^n \left\langle \prod_{j=1}^n \left\{
\sum_{\sigma_j = \pm 1} \sigma_j \frac{\Delta(E + \eta_j)}{\Delta(E + \epsilon_j + i \sigma_j
\kappa_j)} \right\} \right\rangle,\ \ \ \kappa_j \downarrow 0.
\end{equation}
\par
We now derive a semiclassical approximation for $Z_n$.
Using the Gutzwiller's trace formula for chaotic systems\cite{MG}, we can express the trace
of the
resolvent
\begin{equation}
g(E^+) = {\rm Tr}\frac{1}{E^+ - H}
\end{equation}
as a sum over classical periodic orbits $a$
\begin{equation}
\label{traceformula} g(E^+) = {\bar g}(E^+) - \frac{i}{\hbar} \sum_a F_a T_a {\rm e}^{i
S_a(E^+)/\hbar}.
\end{equation}
Here $F_a$ is the stability amplitude (including the Maslov phase), $S_a$ is the classical action and
$T_a = {\rm d}S_a/{\rm d}E$ is the period of $a$. The smoothed part of the trace resolvent is written
as ${\bar g}(E)$. It follows from (\ref{traceformula}) that
\begin{eqnarray}
\label{deltaE} \Delta(E^+) & \propto & {\rm exp}\left( \int^{E^+} g(E') {\rm d}E' \right) \nonumber
\\ & \propto & {\rm exp}\left( - i \pi {\bar N}(E^+) - \sum_a F_a {\rm e}^{i S_a(E^+)/\hbar}
\right).
\end{eqnarray}
Here ${\bar N}(E)$ is the smoothed part of the cumulative energy-level density: it satisfies a
relation ${\bar \rho} = {\rm d} {\bar N}/{\rm d} E$ with the smoothed part of the energy-level
density.
\par
Let us expand the exponential function and write (\ref{deltaE}) as a sum over pseudo-orbits $A$ (a
pseudo-orbit is a set of component periodic orbits):
\begin{equation}
\label{deltaEpseudo}
\Delta(E^+) \propto {\rm e}^{- i \pi {\bar N}(E^+)} \sum_A F_A (-1)^{n_A} \ {\rm e}^{i
S_A(E^+)/\hbar}.
\end{equation}
Here $n_A$ is the number of the component orbits $a$, $S_A$ is the sum of $S_a$ and $F_A$ is the
product of $F_a$. The factor $F_A$ also includes the correction to the sign factor $(-1)^{n_A}$, when
identical orbit copies are contained in $A$. We also find from (\ref{deltaE}) that the inverse of the
spectral determinant is expanded as
\begin{equation}
\label{inversedeltaE} \Delta(E^+)^{-1} \propto {\rm e}^{i \pi {\bar N}(E^+)} \sum_A F_A \ {\rm e}^{i
S_A(E^+)/\hbar}.
\end{equation}
Similar formulas for $\Delta(E^-)$ and $\Delta(E^-)^{-1}$ are obtained by complex conjugation.
\par
However these results do not yet incorporate the unitarity of the quantum-mechanical time evolution,
and thus the fact that the energy levels are real. Berry and Keating argued that the unitarity
requirement of the quantum dynamics leads to an approximation
\begin{equation}
\label{riemann-siegel} \Delta(E) \propto {\rm e}^{- i \pi {\bar N}(E)} \sum_{A \ (T_A < T_H/2)} F_A
(-1)^{n_A} {\rm e}^{i S_A(E)/\hbar} + c.c.
\end{equation}
for a real $E$. This formula is called the Riemann-Siegel lookalike formula\cite{BK00,K02,BK02} after
a similar expression in the theory of the Riemann zeta function.
Here the contributions of ``long'' pseudo-orbits (pseudo-orbits for which the sum $T_A$ of the periods of the component orbits is larger than half the Heisenberg time
$T_H = 2 \pi \hbar {\bar \rho}(E)$) in (\ref{deltaEpseudo})
are replaced by the complex conjugate of the contribution from the shorter pseudo-orbits.
\par
Putting these results into (\ref{Zn}),
we obtain an expression
\begin{eqnarray}
\label{Zn2} & & Z_n = \left( \frac{1}{2 \pi {\bar \rho} i} \right)^n 
\sum_{\substack{\sigma_j = \pm 1 \\ \tau_j = \pm 1}}  
\left\langle {\rm exp}\left[ i \pi \sum_{j=1}^n \{ \sigma_j {\bar N}(E + \epsilon_j)
- \tau_j {\bar N}(E + \eta_j)\} \right] \right. \nonumber \\ & \times & \left. \prod_{j=1}^n \left\{
\sigma_j \sum_{A_j} F^{(\sigma_j)}_{A_j} {\rm e}^{i \sigma_j S_{A_j}(E + \epsilon_j)/\hbar} \sum_{B_j
\ (T_{B_j} < T_H/2)} F^{(\tau_j)}_{B_j} (-1)^{n_{B_j}} {\rm e}^{i \tau_j S_{B_j}(E + \eta_j)/\hbar}
\right\} \right\rangle \nonumber \\
\end{eqnarray}
with $F^{(1)}_A = F_A$ and $F^{(-1)}_A = F^*_A$ (an asterisk means a complex conjugate).
Here the sums over $A_j$ originate from using (\ref{inversedeltaE}) and its complex conjugate
in the denominator of (\ref{Zn}), whereas the sums over $B_j$ result from applying
(\ref{riemann-siegel}) to the numerator. The sums over $\tau_j$ make sure that both summands
in (\ref{riemann-siegel}) are taken into account.
\par
Most of the terms in the sum over $\sigma_j$ and $\tau_j$ vanish when they are averaged over $E$, due
to the highly oscillatory phase factor. Expanding the exponent of the phase factor 
involving ${\bar N}$ in (\ref{Zn2}) as
\begin{eqnarray}
& & {\rm exp}\left[ i \pi \sum_{j=1}^n \{ \sigma_j {\bar N}(E + \epsilon_j) - \tau_j {\bar N}(E +
\eta_j)\} \right] \nonumber \\ & \sim & {\rm exp}\left[ i \pi {\bar N}(E) \sum_{j=1}^n ( \sigma_j -
\tau_j ) + i \pi {\bar \rho}(E) \sum_{j=1}^n ( \sigma_j \epsilon_j - \tau_j \eta_j ) \right],
\end{eqnarray}
we see that such cancellations can be avoided when
\begin{equation}
\label{sigmatau} \sum_{j=1}^n ( \sigma_j - \tau_j  ) = 0
\end{equation}
holds. Hereafter we concentrate on the terms satisfying (\ref{sigmatau}).

\section{The $n$-level correlation functions in the semiclassical limit} 
\setcounter{equation}{0}
\renewcommand{\theequation}{3.\arabic{equation}}

To proceed, we introduce four sets $I$, $J$, $K$ and $L$ so that
\begin{eqnarray}
I&=&\{j|\sigma_j=1\}, \nonumber\\ 
J&=&\{j|\sigma_j=-1\}, \nonumber\\ 
K&=&\{j|\tau_j=1\}, \nonumber\\
L&=&\{j|\tau_j=-1\}.
\end{eqnarray}
It follows from $(\ref{sigmatau})$ that
\begin{equation}
\label{IK} |I| = |K|,
\end{equation}
where $|{\cal M}|$ is the number of the elements when ${\cal M}$ is a set. With this notation the phase factor involving the actions in (\ref{Zn2}) can be written as
\begin{eqnarray}
\label{actions}
& & {\rm exp}\left[ i \sum_{j=1}^n \left\{ \sigma_j S_{A_j}(E + \epsilon_j) + \tau_j S_{B_j}(E +
\eta_j) \right\}/\hbar \right] \nonumber \\ & = & {\rm exp}\left[ i \left\{ \sum_{j \in I} S_{A_j}(E
+ \epsilon_j) - \sum_{k \in J} S_{A_k}(E + \epsilon_k) \right. \right. \nonumber \\ & & \left. \left.
+ \sum_{j \in K} S_{B_j}(E + \eta_j) - \sum_{k \in L} S_{B_k}(E + \eta_k)\right\}/\hbar \right].
\end{eqnarray}
In the semiclassical limit, this phase factor also oscillates rapidly for most choices of pseudo-orbits, meaning
that the corresponding summands in (\ref{Zn2}) will be averaged to zero. In order to find the
dominant contribution, we need to choose the terms with nearly vanishing exponents. For that purpose,
we simply assume that the component orbits in $A_j$, $j \in I$ and $B_j$, $j \in K$ (contributing with a positive sign in (\ref{actions})) are the same as
those in $A_k$, $k \in J$ and $B_k$, $k \in L$ (contributing with a negative sign), neglecting repetitions. We call this scheme the
extended diagonal approximation, as it is a natural extension of Berry's diagonal
approximation\cite{BERRY85}.
\par
Within the diagonal approximation we can now drop the upper limits $T_H/2$ for the 
pseudo-orbits in (\ref{Zn2}). This is possible because each periodic orbit is now a common 
component of two pseudo-orbits. Hence its stability amplitude is coupled with the complex 
conjugate to form the absolute square in (\ref{Zn2}). Weighing orbits with the absolute 
square of their stability amplitude is sufficient to ensure convergence even without an 
upper limit on the sum of periods (for energies with an arbitrarily small imaginary 
part)\cite{KM}.
\par
In the extended diagonal approximation, each pseudo-orbit $A_j$ ($j \in I$)  is a union of the disjoint
sets $A_j \cap A_k$ ($k \in J$) and $A_j \cap B_k$ ($k \in L$). Consequently we find a decomposition
\begin{eqnarray}
\label{dec1} \sum_{A_j} F_{A_j} {\rm e}^{i S_{A_j}(E + \epsilon_j)/\hbar} & = & \prod_{k \in J}
\left( \sum_{A_j \cap A_k} F_{A_j \cap A_k} {\rm e}^{i S_{A_j \cap A_k}(E + \epsilon_j)/\hbar}
\right) \nonumber \\ & \times & \prod_{k \in L} \left( \sum_{A_j \cap B_k} F_{A_j \cap B_k} {\rm
e}^{i S_{A_j \cap B_k}(E + \epsilon_j)/\hbar} \right), \ \ \ j \in I. \nonumber \\
\end{eqnarray}
Analogous decompositions apply to all other pseudo-orbit sums in (\ref{Zn2}). For instance each
pseudo-orbit $A_j$ ($j \in J$) is a union of the disjoint sets $A_j \cap A_k$ ($k \in I$) and $A_j
\cap B_k$ ($k \in K$). It follows that
\begin{eqnarray}
\label{dec2} \sum_{A_j} F^*_{A_j} {\rm e}^{- i S_{A_j}(E + \epsilon_j)/\hbar} & = & \prod_{k \in I}
\left( \sum_{A_j \cap A_k} F^*_{A_j \cap A_k} {\rm e}^{- i S_{A_j \cap A_k}(E + \epsilon_j)/\hbar}
\right) \nonumber \\ & \times & \prod_{k \in K} \left( \sum_{A_j \cap B_k} F^*_{A_j \cap B_k} {\rm
e}^{- i S_{A_j \cap B_k}(E + \epsilon_j)/\hbar} \right), \ \ \ j \in J. \nonumber \\
\end{eqnarray}
Similarly for $j \in K$ and $j \in L$ we obtain the decompositions
\begin{eqnarray}
\label{dec3} & & \sum_{B_j} F_{B_j} (-1)^{n_{B_j}} {\rm e}^{i S_{B_j}(E +
\eta_j)/\hbar} \nonumber \\ & = & \prod_{k \in J} \left( \sum_{B_j \cap A_k} F_{B_j \cap A_k} (-1)^{n_{B_j \cap A_k}} {\rm e}^{i S_{B_j \cap A_k}(E + \eta_j)/\hbar}
\right) \nonumber \\ & \times & \prod_{k \in L} \left( \sum_{B_j \cap B_k} F_{B_j \cap B_k} (-1)^{n_{B_j \cap B_k}} {\rm e}^{i S_{B_j \cap B_k}(E + \eta_j)/\hbar}
\right), \ \ \ j \in K \nonumber \\
\end{eqnarray}
and
\begin{eqnarray}
\label{dec4} & & \sum_{B_j} F^*_{B_j} (-1)^{n_{B_j}} {\rm e}^{-i S_{B_j}(E +
\eta_j)/\hbar} \nonumber \\ & = & \prod_{k \in I} \left( \sum_{B_j \cap A_k} F^*_{B_j \cap A_k} (-1)^{n_{B_j \cap A_k}} {\rm e}^{-i S_{B_j \cap A_k}(E + \eta_j)/\hbar}
\right) \nonumber \\ & \times & \prod_{k \in K} \left( \sum_{B_j \cap B_k} F^*_{B_j \cap B_k} (-1)^{n_{B_j \cap B_k}} {\rm e}^{-i S_{B_j \cap B_k}(E + \eta_j)/\hbar}
\right), \ \ \ j \in L. \nonumber \\
\end{eqnarray}
If we substitute (\ref{dec1}), (\ref{dec2}), (\ref{dec3}) and (\ref{dec4}) into (\ref{Zn2}), we see
that as anticipated the actions almost compensate. The only remaining action difference is 
due to the energy arguments being slightly different. This difference can be 
approximated using expansions of the type
\begin{equation}
\label{actdiff}
S_A(E + \epsilon_j) - S_A(E + \epsilon_k) \sim T_A (\epsilon_j - \epsilon_k).
\end{equation}
Our result thus boils down to
\begin{eqnarray}
\label{znzeta} Z_n & = & \left( \frac{1}{2 \pi {\bar \rho} i} \right)^n 
\sum_{\substack{\sigma_j, \tau_j \\ \left( \sum_{j=1}^n \sigma_j = \sum_{j=1}^n \tau_j \right)}} 
\prod_{j=1}^n \sigma_j {\rm e}^{i \pi
{\bar \rho}(E)(\sigma_j \epsilon_j - \tau_j \eta_j)} \nonumber \\ & \times & \left\langle
\frac{\displaystyle \prod_{j \in I, k \in L} \zeta(-i (\epsilon_j - \eta_k)/\hbar) \prod_{j \in K, k
\in J} \zeta(-i (\eta_j - \epsilon_k)/\hbar)}{ \displaystyle \prod_{j \in I, k \in J} \zeta(-i
(\epsilon_j - \epsilon_k)/\hbar) \prod_{j \in K, k \in L} \zeta(-i (\eta_j - \eta_k)/\hbar)}
\right\rangle, \nonumber \
\end{eqnarray}
where the
sums over intersections of pseudo-orbits (with the remaining action differences (\ref{actdiff}))
were written in terms of the
dynamical zeta function
\begin{eqnarray}
\zeta(s) &=& \sum_{A} |F_A|^2 (-1)^{n_A} {\rm e}^{- s T_A}, 
\nonumber\\ \zeta(s)^{-1} &=& \sum_{A}
|F_A|^2 {\rm e}^{- s T_A}.
\end{eqnarray}
\par
In order to see the asymptotic behavior of $Z_n$ in the semiclassical limit, we introduce the rescaling
\begin{equation}
\label{rescaling}
\epsilon_j \mapsto \frac{\epsilon_j}{2 \pi {\bar \rho}}, \ \ \ j = 1,2,\cdots,n
\end{equation}
and, noting (\ref{rhoE}), utilize the asymptotic formula
\begin{equation}
\label{zeta} \zeta(s) \propto s, \ \ \ s \rightarrow 0,
\end{equation}
which holds for chaotic systems\cite{HAAKE}. In the semiclassical limit $\hbar\to 0$, we now obtain
\begin{eqnarray}
\label{znzetasc} Z_n & = & \left( \frac{1}{2 \pi {\bar \rho} i} \right)^n 
\sum_{\substack{\sigma_j, \tau_j \\ \left( \sum_{j=1}^n \sigma_j = \sum_{j=1}^n \tau_j \right)}} 
(-1)^{|L|} {\rm e}^{i \left( \sum_{j \in I}
\epsilon_j - \sum_{j \in J} \epsilon_j  - \sum_{j \in K} \eta_j + \sum_{j \in L} \eta_j\right)/2}
\nonumber\\
& \times & \frac{\displaystyle \prod_{\substack{j \in I \\ k \in L}} (\epsilon_j - \eta_k) \prod_{\substack{j \in K \\ k \in J}}
(\eta_j - \epsilon_k)}{ \displaystyle \prod_{\substack{j \in I \\ k \in J}} (\epsilon_j - \epsilon_k) 
\prod_{\substack{j \in K \\ k \in L}} (\eta_j - \eta_k)}.
\end{eqnarray}
\par
To compare this result to the random-matrix expression in \cite{CS08}, 
it is helpful to adopt a slightly different notation. The sum over
all choices for the sign factors $\tau_j=\pm 1$, $j=1,2,\ldots,n$ is equivalent to summation over all ways to
write the set $\{1,2,\cdots,n\}$ as a direct sum of two subsets $K$ and $L$. The direct sum of disjoint 
sets $K$ and $L$ is defined as the union $K\cup L$ and denoted by $K+L$. The corresponding arguments 
$i\epsilon_j$, $i\eta_j$ in (\ref{znzetasc}) then form sets
\begin{eqnarray}
\label{abcd} &&A = \{ i \epsilon_j| j \in K \}, \ \ \ B = \{ - i \epsilon_j| j \in L \}, 
\nonumber\\ &&C = \{ i \eta_j| j \in K \}, \ \ \ D = \{ - i \eta_j| j \in L \}.
\end{eqnarray}
Moreover the sum over signs $\sigma_j=\pm 1$ determining the sets $I$ and $J$
can be replaced by a sum over subsets
\begin{eqnarray}
S&=&\{i\epsilon_j|j\in J\cap K\}\subset A, 
\nonumber\\ T&=&\{-i\epsilon_j|j\in I\cap L\}\subset B.
\end{eqnarray}
Indeed, if we define
\begin{equation}
\label{sbtb} {\bar S} = A - S, \ \ \ {\bar T} = B - T,
\end{equation}
and ${\cal M}^- = \{ - \alpha | \alpha \in {\cal M}\}$ when ${\cal M}$ is a set, the sets of energy
increments corresponding to $I$ and $J$ can be expressed through $S$ and $T$ as
\begin{eqnarray}
{\bar S} + T^- &=&\{i\epsilon_j|j\in I\}, \nonumber \\ 
{\bar T} + S^-  &=&\{-i\epsilon_j|j\in J\}.
\end{eqnarray}
With these definitions, when we apply the rescaling (\ref{rescaling}) 
to (\ref{RnZn}), Eq.(\ref{znzetasc}) yields 
\begin{equation}
\label{Rn} R_n(\epsilon_1,\cdots,\epsilon_n) = \sum_{K+L = \{1,2,\cdots,n\}} q(A;B),
\end{equation}
where
\begin{equation}
\label{q}
q(A;B) = \prod_{\alpha \in A, \beta \in B} \left. \frac{{\rm \partial}}{{\rm \partial} \alpha}
\frac{{\rm \partial}}{{\rm \partial} \beta} r(A,B;C,D) \right|_{\boldsymbol\eta=\boldsymbol\epsilon}
\end{equation}
and $r(A,B;C,D)$ is defined as
\begin{eqnarray}
r(A,B;C,D) & = & \sum_{\substack{S \subset A, T \subset B \\ (|S| = |T|)}} {\rm exp}\left\{ \frac{1}{2}
\left(\sum_{\alpha \in {\bar S} + T^-} \alpha + \sum_{\beta \in {\bar T} + S^-} \beta - \sum_{\gamma
\in C} \gamma - \sum_{\delta \in D} \delta \right) \right\}  \nonumber \\ & \times & z({\bar S} +
T^-, {\bar T} + S^-; C,D)
\end{eqnarray}
with
\begin{equation}
\label{zwxyz} z({\cal W},{\cal X};{\cal Y},{\cal Z}) 
= \frac{\displaystyle \prod_{\substack{\alpha \in {\cal W} \\ \delta \in {\cal Z}}} (\alpha + \delta)
\prod_{\substack{\beta \in {\cal X} \\ \gamma \in {\cal Y}}} (\beta + \gamma)}{
\displaystyle \prod_{\substack{\alpha \in {\cal W} \\ \beta \in {\cal X}}} (\alpha + \beta) 
\prod_{\substack{\gamma \in {\cal Y} \\ \delta \in {\cal Z}}} (\gamma + \delta)}.
\end{equation}
\par
For example, the $2$-level correlation function is calculated as
\begin{eqnarray}
\label{R2} R_2(\epsilon_1,\epsilon_2) & = & q(\emptyset;\{-i \epsilon_1,- i \epsilon_2\}) + q(\{i
\epsilon_1 \}; \{-i \epsilon_2\}) \nonumber \\ & + & q(\{i \epsilon_2 \}; \{-i \epsilon_1\}) + q(\{i
\epsilon_1,i \epsilon_2\};\emptyset),
\end{eqnarray}
where
\begin{eqnarray}
\label{R2-1} q(\emptyset;\{-i \epsilon_1,- i \epsilon_2\}) & = & - \frac{\partial^2}{\partial
\epsilon_1 \partial \epsilon_2} r(\emptyset,\{- i \epsilon_1, - i \epsilon_2\};\emptyset, \{- i
\eta_1,-i \eta_2\}) \nonumber \\ & = & \left. - \frac{\partial^2}{\partial \epsilon_1 \partial
\epsilon_2} {\rm e}^{-i(\epsilon_1 + \epsilon_2 - \eta_1 - \eta_2)/2} \right|_{\boldsymbol\eta =
\boldsymbol\epsilon} \nonumber \\ & = & \frac{1}{4}, \\  
\label{R2-2} q(\{i \epsilon_1\};\{- i \epsilon_2\}) & = & \frac{\partial^2}{\partial \epsilon_1
\partial \epsilon_2} r(\{i \epsilon_1\},\{- i \epsilon_2\};\{i \eta_1\}, \{- i \eta_2\}) \nonumber \\
& = & \frac{\partial^2}{\partial \epsilon_1 \partial \epsilon_2} \left\{ {\rm e}^{i(\epsilon_1 -
\epsilon_2 - \eta_1 + \eta_2)/2} \frac{(\epsilon_1 - \eta_2)(\eta_1 - \epsilon_2)}{(\epsilon_1 -
\epsilon_2) (\eta_1 - \eta_2)} \right. \nonumber \\ && +  \left. \left. {\rm e}^{i(\epsilon_2 -
\epsilon_1 - \eta_1 + \eta_2)/2} \frac{(\epsilon_2 - \eta_2)(\eta_1 - \epsilon_1)}{(\epsilon_2 -
\epsilon_1) (\eta_1 - \eta_2)} \right\} \right|_{\boldsymbol\eta = \boldsymbol\epsilon}
 \nonumber \\
& = & \frac{1}{4} - \frac{1 - {\rm e}^{-i (\epsilon_1 - \epsilon_2)}}{ (\epsilon_1 -
\epsilon_2)^2}, \\ 
\label{R2-3} q(\{i \epsilon_2\};\{- i \epsilon_1\})
 &=& \frac{1}{4} - \frac{1 - {\rm e}^{i (\epsilon_1 - \epsilon_2)}}{
(\epsilon_1 - \epsilon_2)^2}, \\ 
\label{R2-4} q(\{i \epsilon_1,i \epsilon_2\};\emptyset) &=&
\frac{1}{4}.
\end{eqnarray}
Putting (\ref{R2-1}), (\ref{R2-2}), (\ref{R2-3}) and (\ref{R2-4}) into (\ref{R2}), we can readily
find a compact expression
\begin{equation}
R_2(\epsilon_1,\epsilon_2) = 1 - \left[ \frac{\sin\{ (\epsilon_1 - \epsilon_2)/2 \}}{(\epsilon_1 -
\epsilon_2)/2} \right]^2.
\end{equation}

\section{Determinant expressions}
\setcounter{equation}{0}
\renewcommand{\theequation}{4.\arabic{equation}}

In \cite{CS08}, Conrey and Snaith analyzed the Circular Unitary Ensemble (CUE) of random matrices.
Based on the Ratios Theorem\cite{CFZ,CFS} (see also \cite{FS,SF}) 
on the characteristic polynomials, they established the formulas
\begin{equation}
\label{RCUEn} R^{({\rm CUE})}_n(\epsilon_1,\cdots,\epsilon_n) = \sum_{K+L+M = \{1,2,\cdots,n\}}
{\tilde q}(A;B)
\end{equation}
for the scaled $n$-eigenparameter correlation functions $R^{({\rm CUE})}_n$. Here the union of the
three disjoint sets $K$, $L$ and $M$ is $\{1,2,\cdots,n \}$ and
\begin{equation}
{\tilde q}(A;B) = \prod_{\alpha \in A, \beta \in B} \left. \frac{{\rm \partial}}{{\rm \partial}
\alpha} \frac{{\rm \partial}}{{\rm \partial} \beta} {\tilde r}(A,B;C,D) \right|_{\boldsymbol\eta =
\boldsymbol\epsilon}
\end{equation}
with
\begin{equation}
{\tilde r}(A,B;C,D) = \sum_{\substack{S \subset A, T \subset B \\ (|S| = |T|)}} {\rm exp}\left\{ - \sum_{\alpha
\in S} \alpha - \sum_{\beta \in T} \beta \right\} z({\bar S} + T^-, {\bar T} + S^-; C,D).
\end{equation}
The definitions of $A$, $B$, $C$, $D$, ${\bar S}$, ${\bar T}$ and 
$z({\cal W},{\cal X};{\cal Y},{\cal Z})$ are the same as before 
((\ref{abcd}), (\ref{sbtb}) and (\ref{zwxyz})).
\par
On the other hand, it is well known that the scaled $n$-eigenparameter correlation functions of the
CUE have determinant expressions\cite{DY1962}:
\begin{equation}
R^{({\rm CUE})}_n(\epsilon_1,\cdots,\epsilon_n) = \det\left[ \frac{\sin\{(\epsilon_j -
\epsilon_k)/2\}}{(\epsilon_j - \epsilon_k)/2} \right]_{j,k=1,2,\cdots,n}.
\end{equation}
Therefore, in order to verify the same determinant expressions for the scaled semiclassical $n$-level
correlation functions (\ref{Rn}), it is sufficient to prove that (\ref{Rn}) is identical to
(\ref{RCUEn}).
\par
To do this, we note that the derivatives $\frac{\partial}{\partial\alpha}$ ($\alpha\in A$),
$\frac{\partial}{\partial\beta}$ ($\beta\in B$) in (\ref{q}) may act either on the phase factor
\begin{equation}
\phi = {\rm exp}\left\{ \frac{1}{2} \left(\sum_{\alpha \in {\bar S} + T^-} \alpha + \sum_{\beta \in
{\bar T} + S^-} \beta - \sum_{\gamma \in C} \gamma - \sum_{\delta \in D} \delta \right) \right\}
\end{equation}
or on $z({\bar S} + T^-, {\bar T} + S^-; C,D)$. We can thus sum over all ways to split $A$ into two
disjoint subsets $A_1$ and $A_2$, and then let the derivatives with respect to the elements of $A_1$ 
act on $\phi$ whereas the derivatives with respect to the elements of $A_2$ act on $z$. 
The corresponding sets of indices in $K$
are denoted by $K_1$ and $K_2$. Analogously $B$ is divided into subsets $B_1$ (with derivatives
acting on $\phi$) and $B_2$ (with derivatives acting on $z$), and the corresponding sets of indices
in $L$ are denoted by $L_1$ and $L_2$. This yields
\begin{eqnarray}
\label{Rnrw} 
& & R_n(\epsilon_1,\cdots,\epsilon_n) \nonumber \\ & = & \sum_{K+L = \{1,2,\cdots,n\}} 
\sum_{\substack{S \subset A, T \subset B\ (|S| = |T|) \\ A_1+A_2=A,B_1+B_2=B}} \left( \prod_{\alpha \in A_1, \beta \in B_1}
\frac{{\rm \partial}}{{\rm \partial} \alpha} \frac{{\rm \partial}}{{\rm \partial} \beta} \phi \right)
\nonumber \\ & \times & \left. \left\{ \prod_{\alpha \in A_2, \beta \in B_2} \frac{{\rm
\partial}}{{\rm \partial} \alpha} \frac{{\rm \partial}}{{\rm \partial} \beta} z({\bar S} + T^-, {\bar
T} + S^-; C,D) \right\} \right|_{\boldsymbol\eta = \boldsymbol\epsilon}.
\end{eqnarray}
Now it is important that the derivatives
\begin{equation}
\left. \prod_{\alpha \in A_2, \beta \in B_2} \frac{{\rm \partial}}{{\rm \partial} \alpha} \frac{{\rm
\partial}}{{\rm \partial} \beta} z({\bar S} + T^-, {\bar T} + S^-; C,D) \right|_{\boldsymbol\eta =
\boldsymbol\epsilon}
\end{equation}
in (\ref{Rnrw}) are nonzero only if 
\begin{equation}
\label{sa2tb2}
S \subset A_2 \ {\rm  and} \ T \subset B_2
\end{equation}
hold. This is because, for each element $i\epsilon_j\in S$,
$-i\epsilon_j$ is included in $S^-\subset {\bar T} + S^-$ and the 
corresponding element $i\eta_j$ is included in $C$ so that 
$z({\bar S} + T^-,{\bar T} + S^-;C,D)$ defined in (\ref{zwxyz}) 
contains a factor $-i\epsilon_j+i\eta_j$ vanishing for $\boldsymbol\eta 
= \boldsymbol\epsilon$. Nonzero contributions arise only 
if all such terms are eliminated by differentiating $z$ with respect to 
all elements $i\epsilon_j\in S$. We thus need to have 
$S\subset A_2$. Analogous reasoning leads to $T\subset B_2$.
\par
Now let us consider the phase factor $\phi$. Each derivative of $\phi$ with 
respect to the elements $\alpha\in A_1\subset {\bar S}$ and $\beta\in B_1\subset 
{\bar T}$ leads to a factor $\frac{1}{2}$. If we subsequently identify 
$\boldsymbol\eta=\boldsymbol\epsilon$, the exponent of $\phi$ turns into
\begin{equation}
\frac{1}{2}\left(\sum_{\alpha \in {\bar S}+T^{-}}\alpha+\sum_{\beta \in 
{\bar T}+S^{-}}\beta
-\sum_{\gamma\in C=A=S+ {\bar S}}\gamma-\sum_{\delta \in 
D=B=T+ {\bar T}}\delta\right)
=-\sum_{\alpha\in S}\alpha-\sum_{\beta\in T}\beta\,.
\end{equation}
We thus obtain
\begin{eqnarray}
\label{Rnsimp}
& &R_n(\epsilon_1,\cdots,\epsilon_n) =  \sum_{K+L = \{1,2,\cdots,n\}} 
\sum_{\substack{S \subset A_2 \subset A, T \subset B_2 \subset B
\\ (|S| = |T|)}} \frac{1}{ 2^{|A_1|+|B_1|}} \nonumber \\ 
& \times & {\rm e}^{- \sum_{\alpha \in S} \alpha -
\sum_{\beta \in T} \beta} \prod_{\alpha \in A_2, \beta \in B_2} \left. \frac{{\rm \partial}}{{\rm
\partial} \alpha} \frac{{\rm \partial}}{{\rm \partial} \beta} z({\bar S} + T^-, {\bar T} + S^-; C,D)
\right|_{\boldsymbol\eta = \boldsymbol\epsilon}. \nonumber \\ 
\end{eqnarray}
\par
Eq.(\ref{Rnsimp}) can be further simplified, if we use the fact that
the summands in (\ref{Rnsimp}) do not depend on how the 
elements of $A_1+B_1$ are distributed among $A_1$ and $B_1$. 
In particular, we have 
\begin{eqnarray}
\label{indep}
&&\prod_{\alpha \in A_2, \beta \in B_2} \left. \frac{{\rm \partial}}{{\rm
\partial} \alpha} \frac{{\rm \partial}}{{\rm \partial} \beta} z({\bar S} + T^-, {\bar T} + S^-; C,D)
\right|_{\boldsymbol\eta = \boldsymbol\epsilon}\nonumber\\
&=&
\prod_{\alpha \in A_2, \beta \in B_2} \left. \frac{{\rm \partial}}{{\rm
\partial} \alpha} \frac{{\rm \partial}}{{\rm \partial} \beta} z({\bar S}\cap A_2 + T^-, {\bar T}\cap B_2 + S^-; C_2,D_2)
\right|_{\boldsymbol\eta = \boldsymbol\epsilon},
\end{eqnarray}
where
\begin{eqnarray}
C_2&=&\{i\eta_j|j\in K_2\}\subset C, 
\nonumber\\
D_2&=&\{-i\eta_j|j\in L_2\}\subset D
\end{eqnarray}
and all sets on the r.h.s. of (\ref{indep}) can be shown to exclude $A_1$ and $B_1$. 
\par
Let us prove (\ref{indep}). It follows from (\ref{sa2tb2}) that 
${\bar S}={\bar S}\cap A_1+{\bar S}\cap A_2 =A_1+{\bar S}\cap A_2$ 
and ${\bar T}={\bar T}\cap B_1+{\bar T}\cap B_2 = B_1+{\bar T}\cap B_2$. 
Then, using $C_1 = C - C_2$ and $D_1 = D - D_2$, we obtain 
\begin{eqnarray}
\label{zfact}
& & z({\bar S} + T^-, {\bar T} + S^-; C,D)
 = \frac{\displaystyle \prod_{
\substack{\alpha \in  A_1+ \bar S\cap A_2 + T^- \\ \delta \in D_1 + D_2}} (\alpha + \delta) 
\prod_{\substack{\beta \in B_1 + {\bar T} \cap B_2 + S^- \\ \gamma \in C_1 + C_2}} (\beta + \gamma)}{
\displaystyle
\prod_{\substack{\alpha \in  A_1 + {\bar S} \cap A_2 + T^- \\ \beta \in B_1 + {\bar T} \cap B_2 + S^-}} 
(\alpha + \beta) 
\prod_{\substack{\gamma \in C_1 + C_2 \\ \delta \in D_1 + D_2}} (\gamma + \delta)} \nonumber \\ & = &
W_1 W_2 z({\bar S} \cap A_2 + T^-, {\bar T} \cap B_2 + S^-;C_2,D_2),
\end{eqnarray}
where 
\begin{equation}
W_1 = \frac{\displaystyle \prod_{\substack{\alpha \in  A_1 \\ \delta \in D}} (\alpha + \delta) 
\prod_{\substack{\beta \in B_1 \\ \gamma \in C}} (\beta + \gamma)}{
\displaystyle 
\prod_{\substack{\alpha \in  A_1 \\ \beta \in B_1}} (\alpha + \beta) 
\prod_{\substack{\gamma \in C_1 \\ \delta \in D_1}} (\gamma + \delta) 
\prod_{\substack{\gamma \in C_1 \\ \delta \in D_2}} (\gamma + \delta) 
\prod_{\substack{\gamma \in C_2 \\ \delta \in D_1}} (\gamma + \delta)}
\end{equation}
and
\begin{eqnarray}
\label{W2}
& & W_2 = \frac{\displaystyle 
\prod_{\substack{\alpha \in  {\bar S} \cap A_2 + T^- \\ \delta \in D_1}} (\alpha + \delta) 
\prod_{\substack{\beta \in {\bar T} \cap B_2 + S^- \\ \gamma \in C_1}} (\beta + \gamma)}{
\displaystyle 
\prod_{\substack{\alpha \in  A_1 \\ \beta \in {\bar T} \cap B_2 + S^-}} (\alpha + \beta)
\prod_{\substack{\alpha \in  {\bar S} \cap A_2 + T^- \\ \beta \in B_1}} (\alpha + \beta)} 
\nonumber \\ & = & \displaystyle 
\prod_{ \alpha \in  {\bar S} \cap A_2 + T^-} \prod_{j \in L_1} 
\left( 1 + i \frac{\epsilon_j - \eta_j}{\alpha - i \epsilon_j} \right) 
\prod_{\beta \in {\bar T} \cap B_2 + S^-}
\prod_{j \in K_1} \left( 1 + i \frac{\eta_j - \epsilon_j}{\beta + i \epsilon_j} \right). 
\nonumber \\ 
\end{eqnarray}
We can see that $W_1$ is independent of the elements of $A_2 \cup B_2$ and
\begin{equation}
\label{w2}
\left. \frac{\partial}{\partial \alpha} W_2 \right|_{ \boldsymbol\eta = \boldsymbol\epsilon} = 0
\end{equation}
for $\alpha \in A_2 \cup B_2$. (To check (\ref{w2}), note that only the $\alpha$ 
and $\beta$ in the last line of (\ref{W2}) may belong to $A_2$ or $B_2$. If we take
derivatives with respect to any of these variables, one factor in the product turns into
$-i\frac{\epsilon_j-\eta_j}{(\alpha-i\epsilon_j)^2}$ or $-i\frac{\eta_j-\epsilon_j}{(\beta+i\epsilon_j)^2}$ and vanishes after setting $\boldsymbol\eta
=\boldsymbol\epsilon$.) Therefore nonvanishing contributions arise only
if the derivatives of (\ref{zfact}) with respect to the elements of 
$A_2 \cup B_2$ act on $z({\bar S} \cap A_2 + T^-, {\bar T} \cap B_2 + S^-;C_2,D_2)$. Then $W_1$ and 
$W_2$ can be replaced by their special values 
\begin{equation}
\left. W_1 \right|_{\boldsymbol\eta = \boldsymbol\epsilon} = \left. W_2 \right|_{\boldsymbol\eta =
\boldsymbol\epsilon} = 1
\end{equation}
at $\boldsymbol\eta =\boldsymbol\epsilon$. 
\par
Thus Eq.(\ref{indep}) is proven, and it indeed becomes irrelevant on 
how the elements of $A_1+B_1$ are distributed among the two subsets, or on how the corresponding indices in $M\equiv K_1+L_1$ are distributed among $K_1$ and $L_1$. We can thus stop to discriminate between $K_1$ and $L_1$.
This means that we trade the sums over $K,L$ and over $A_2,B_2$ in (\ref{Rnsimp}) for one over $K_2+L_2+M=\{1,2,\ldots,n\}$. Since there are $2^{|M|}$ ways to
divide $M$ into $K_1$ and $L_1$ we then need to multiply with $2^{|M|}$ which
cancels the factor $\frac{1}{2^{|A_1|+|B_1|}}=\frac{1}{2^{|M|}}$. 
\par
The $n$-level correlation functions can thus be written as
\begin{eqnarray}
& & R_n(\epsilon_1,\cdots,\epsilon_n) = \sum_{K_2+L_2+M = \{1,2,\cdots,n\}} 
\sum_{\substack{S \subset A_2, T \subset B_2 \\ (|S| = |T|)}} 
{\rm e}^{- \sum_{\alpha \in S} \alpha - \sum_{\beta \in T} \beta}
\nonumber \\ & \times & \prod_{\alpha \in A_2, \beta \in B_2} \left. \frac{{\rm \partial}}{{\rm
\partial} \alpha} \frac{{\rm \partial}}{{\rm \partial} \beta} z({\bar S} \cap A_2 + T^-, {\bar T} \cap
B_2+ S^-; C_2,D_2) \right|_{\boldsymbol\eta = \boldsymbol\epsilon}, \nonumber \\
\end{eqnarray}
which is identical to the CUE $n$-eigenparameter correlation functions (\ref{RCUEn}).
The sets $K_2,L_2,A_2,B_2$ correspond to $K,L,A,B$ in (\ref{RCUEn}). The determinant expressions
\begin{equation}
 R_n(\epsilon_1,\cdots,\epsilon_n) =
\det\left[ \frac{\sin\{(\epsilon_j - \epsilon_k)/2\}}{(\epsilon_j - \epsilon_k)/2}
\right]_{j,k=1,2,\cdots,n}
\end{equation}
for the $n$-level correlation functions are thus verified.

\section{Discussion}

In this paper the spectral correlation functions of chaotic quantum systems with broken time-reversal
symmetry were investigated. Using the Riemann-Siegel lookalike formula and
an extended version of Berry's diagonal approximation, we semiclassically evaluated the $n$-level
correlation functions identical to the $n \times n$ determinantal predictions of random matrix
theory.
\par
Although the semiclassical results are exactly in agreement with the random matrix predictions, we
admit that there are higher order nonzero terms in the semiclassical diagrammatic expansion, which
are neglected in the extended diagonal approximation. The exact agreement suggests that those terms
mutually cancel each other. Such a cancellation was diagrammatically verified in \cite{H07,M04, M05,SM} 
for the $2$-level correlation function. It should also be verified for the
general $n$-level correlation functions in future works.
\par
As noted in Introduction, if the quantum system is symmetric under time-reversal, it belongs to a
different universality class. In this case, the semiclassical argument becomes more difficult,
because the higher order terms in the diagrammatic expansion are more involved and give a net
contribution\cite{H07,M04,M05,SM}. The random matrix predictions are derived from the corresponding
Circular Orthogonal Ensemble (COE) (or Gaussian Orthogonal Ensemble (GOE)), and the $n$-level
correlation functions have $2 n \times 2 n$ Pfaffian forms\cite{DY1970}. It would be interesting if
difficulties are overcome and one is able to see how Pfaffian forms semiclassically appear.

\section*{Acknowledgements}

This work was partially supported by Japan Society for the Promotion of Science (KAKENHI 20540372).
The authors are grateful to Dr. Martin Sieber, Dr. Nina C. Snaith and 
Prof. Jonathan P. Keating for valuable discussions.

\end{document}